\documentclass[prl,twocolumn,showpacs,amsmath,amssymb]{revtex4}

\usepackage{graphicx}
\usepackage{dcolumn}
\usepackage{bm}

\begin{document}


\title {Interband Recombination Dynamics in\\
Resonantly-Excited Single-Walled Carbon Nanotubes}
\author{G. N. Ostojic, S. Zaric, J. Kono}
 \thanks{Author to whom correspondence should be addressed}
 \email{kono@rice.edu}
\affiliation{Department of Electrical and Computer Engineering,
Rice Quantum Institute, and Center for Nanoscale Science and Technology,
Rice University, Houston, Texas 77005}
\author{M. S. Strano, V. C. Moore, R. H. Hauge, R. E. Smalley}
\affiliation{Department of Chemistry,
Rice Quantum Institute, and Center for Nanoscale Science and Technology,
Rice University, Houston, Texas 77005}

\date{\today}

\begin{abstract}

Wavelength-dependent pump-probe spectroscopy of
micelle-suspended single-walled carbon nanotubes
reveals two-component dynamics.
The slow component (5-20 ps), which has
not been observed previously,
is resonantly enhanced whenever the pump photon energy
coincides with an absorption peak and we attribute it to
interband carrier recombination,
whereas we interpret the always-present
fast component (0.3-1.2 ps) as intraband carrier relaxation in
non-resonantly excited nanotubes.
The slow component decreases drastically
with decreasing pH (or increasing H$^+$ doping), especially in
large-diameter tubes.  This can be explained as a consequence of
the disappearance of absorption peaks at high doping
due to the entrance of the Fermi energy into the
valence band, i.e., a 1-D manifestation of the Burstein-Moss effect.

\end{abstract}

\maketitle

Optical properties of one-dimensional (1-D) systems have been the subject
of numerous theoretical studies for many years \cite{loudon}.
Various 1-D systems have been investigated experimentally,
but the distinct characters of 1-D excitons
have not been revealed unambiguously.
Single-walled carbon nanotubes (SWNTs)
\cite{dresselhaus01} provide a viable alternative for exploring 1-D
exciton physics, and, in addition, they are expected to show a new class of
optical phenomena that arise from their unique tubular structure.
Linear and nonlinear optical coefficients are expected to be
diameter- and chirality-dependent, and a magnetic field applied parallel to the tubes
is expected to induce non-intuitive modifications on
their electronic, magnetic, and optical properties via the Aharonov-Bohm phase
\cite{ajikiJPSJ93}.
Furthermore, high-order harmonic generation is expected to be extremely
selective \cite{alonPRL00}.

However, these theoretical predictions are mostly unexplored experimentally.
The main reason is that SWNTs easily
form bundles (or `ropes'), consisting of different tube types,
due to their strong van der Waals forces.
This results in significant broadening,
smearing out any chirality-dependent
features \cite{katauraSM99}.
Very recently, a new technique for
producing individually-suspended SWNTs in solutions has been reported
\cite{Science1}.  It consists of separating tubes by vigorous sonication,
followed by micelle wrapping of individual tubes to prevent
bundling.  These samples have revealed, for the first time, a number of
clearly observable peaks in linear absorption and photoluminescence (PL) spectra,
corresponding to interband transitions in different types of tubes.
Subsequent photoluminescence-excitation (PLE) spectroscopy studies successfully
provided detailed peak assignments \cite{Science2,Weisman03NL}.

\begin{figure}
\begin{center}
\includegraphics [scale=0.64] {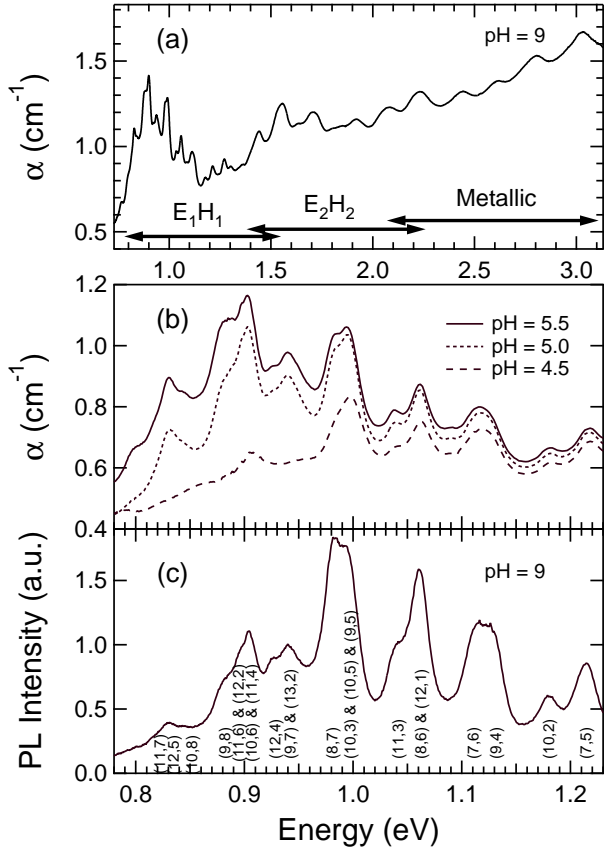}
\caption{(a) Linear absorption spectrum for first (E$_1$H$_1$) and
second (E$_2$H$_2$) subband transitions in semiconducting SWNTs and first subband
transitions in metallic SWNTs.
(b) Linear absorption spectra in the E$_1$H$_1$ range for pH = 5.5., 5.0, and 4.5.
(c) Photoluminescence spectrum taken with an excitation wavelength
of 632 nm.}
\label{linear}
\end{center}
\end{figure}

Here we report results of wavelength-dependent
pump-probe spectroscopy on such micelle-suspended and
chirality-assigned SWNTs.  Unlike the bundled nanotubes used in
previous pump-probe experiments \cite{PP1,PP2,PP3,PP4},
our samples show
distinct interband absorption and PL peaks (see Fig.~\ref{linear}).
By scanning the photon energy in a wide range, we were
able to identify different types of relaxation processes.
Specifically, we discovered two distinct relaxation regimes: fast (0.3-1.5 ps)
and slow (5-20 ps).  The slow component, which has not been observed
previously, is {\em resonantly enhanced} whenever the pump photon energy
coincides with an absorption peak, and we ascribe it
to interband carrier recombination in resonantly-excited nanotubes.
The fast component is always present, and we attribute it to intraband carrier
relaxation in non-resonantly-excited nanotubes.
Furthermore, the slow component decreases drastically
with decreasing pH (or increasing H$^+$ doping), especially in
large-diameter tubes.  We interpret this behavior as a consequence of
the Burstein-Moss effect \cite{BM}.

The SWNTs studied in the present work were dispersed in aqueous sodium dodecyl
sulfate (SDS) surfactant, sonicated, and centrifuged.
The sample preparation method was described in \cite{Science1,Science2}.
Wavelength-dependent degenerate pump-probe measurements were performed
using $\sim$150 fs pulses from an optical parametric amplifier (OPA)
pumped by a chirped pulse amplifier (Clark-MXR CPA2010).  The
low pulse repetition rate (1 kHz) was ideal for
reducing any thermal effects
while keeping the pump fluence high.  To record small photoinduced changes in
probe transmission, we synchronously chopped the pump beam at 500 Hz and
measured the transmission with ($T$) and without ($T_{0}$)
the pump using two different gates of a box-car integrator.

Figure \ref{linear}(a) shows an absorption spectrum,
exhibiting three absorption bands: i) first subband (E$_1$H$_1$) transitions
in semiconducting tubes, ii) second subband (E$_2$H$_2$) transitions
in semiconducting tubes,
and iii) lowest-energy transitions in metallic tubes.
Figures \ref{linear}(b) and \ref{linear}(c) show absorption and PL in the
E$_1$H$_1$ range.  Three traces in (b) correspond to different pH values.
It is seen that absorption peaks at long wavelengths diminish with
decreasing pH, similar to a recent report \cite{StranoetAl03JPCB}.
\begin{figure}
\begin{center}
\includegraphics [scale=0.5] {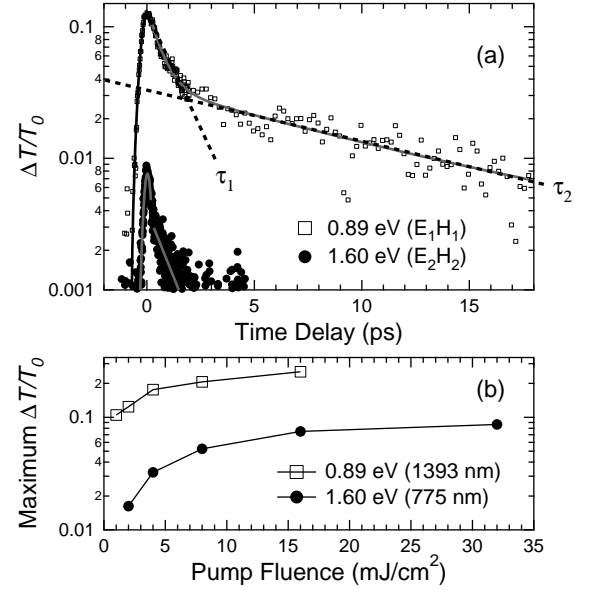}
\caption{(a) Pump-probe data at two wavelengths,
corresponding to first and second subband transitions.
(b) Maximum transmission change vs.~pump fluence.}
\label{typical}
\end{center}
\end{figure}

Figure \ref{typical}(a) shows typical pump-probe data.  Two traces were
taken at 0.89 eV (1393 nm) and 1.60 eV (775 nm),
corresponding to E$_1$H$_1$ and
E$_2$H$_2$ transitions, respectively.
Both show a positive change in
transmission, i.e., photoinduced bleaching, consistent with
state filling. An exponential fit reveals a fast, single decay time of 770 fs for the
E$_2$H$_2$ transition, consistent with
intraband relaxation towards the band edge \cite{PP4}.
On the contrary, data in the range of
E$_1$H$_1$ transitions exhibit double-exponential decays.  The major
decay happens in the first picosecond (with
decay time $\tau_1$), which is followed by
slower relaxation (with $\tau_2$).  For the particular data in
Fig.~\ref{typical}(a),
we obtain $\tau_2$ $\approx$ 10 ps.  This long
decay time has not been reported previously in either photo-emission \cite{PP1}
or pump-probe studies \cite{PP2,PP3,PP4}.

For both first and second subband transitions, the pump fluence
dependence of the maximum value of $\Delta T/T_0$ reveals clear
saturation at high fluences, as shown in Fig.~\ref{typical}(b).
This implies
that, in the saturation regime, most of the carrier states are filled and
thus the sample absorption is nearly completely quenched.  A careful
analysis of the differential transmission decays for
0.89 eV showed that relaxation
dynamics are not dependent on the pump fluence, including the saturation
regime.
\begin{figure}
\begin{center}
\includegraphics [scale=0.55] {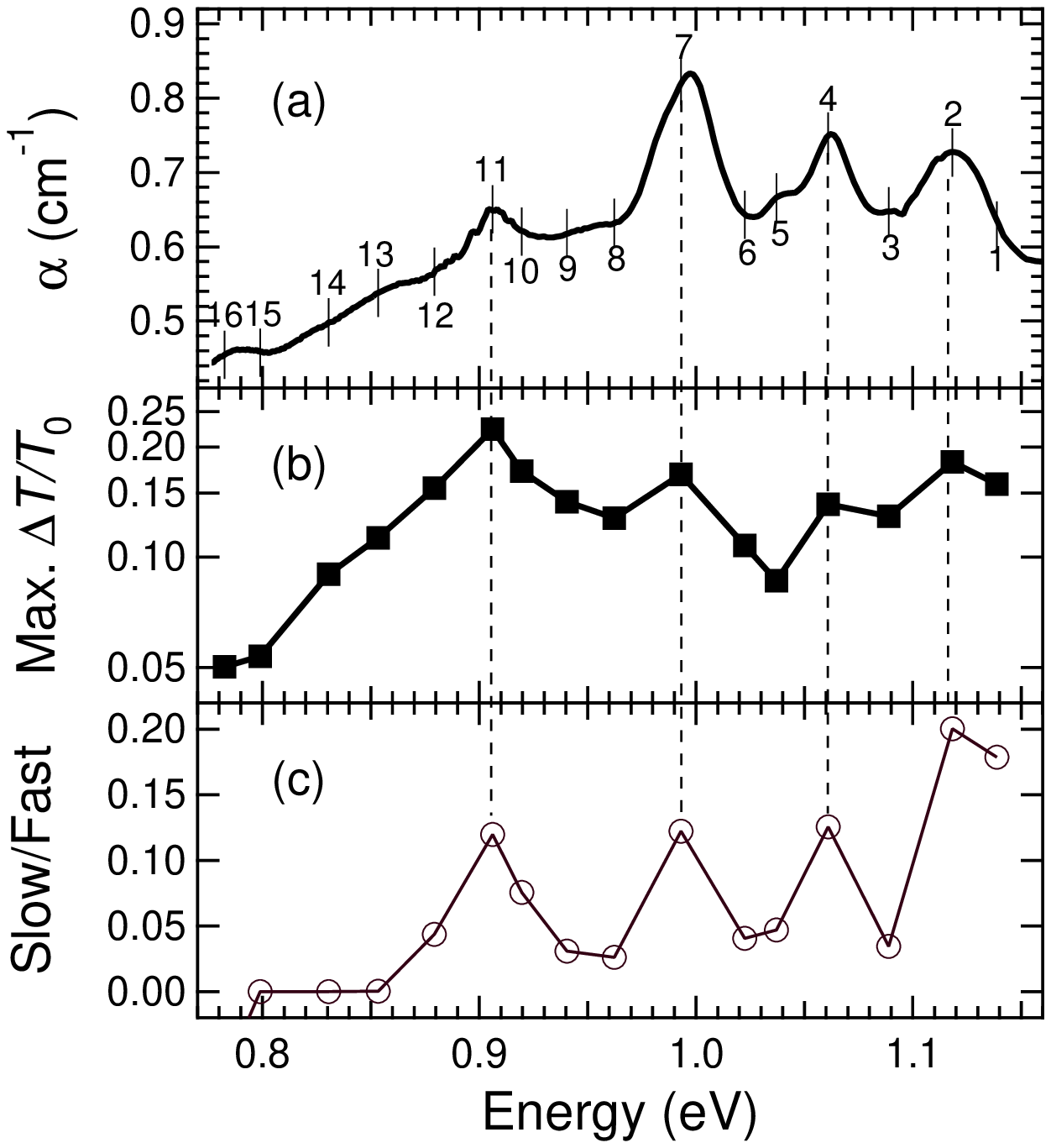}
\includegraphics [scale=0.57] {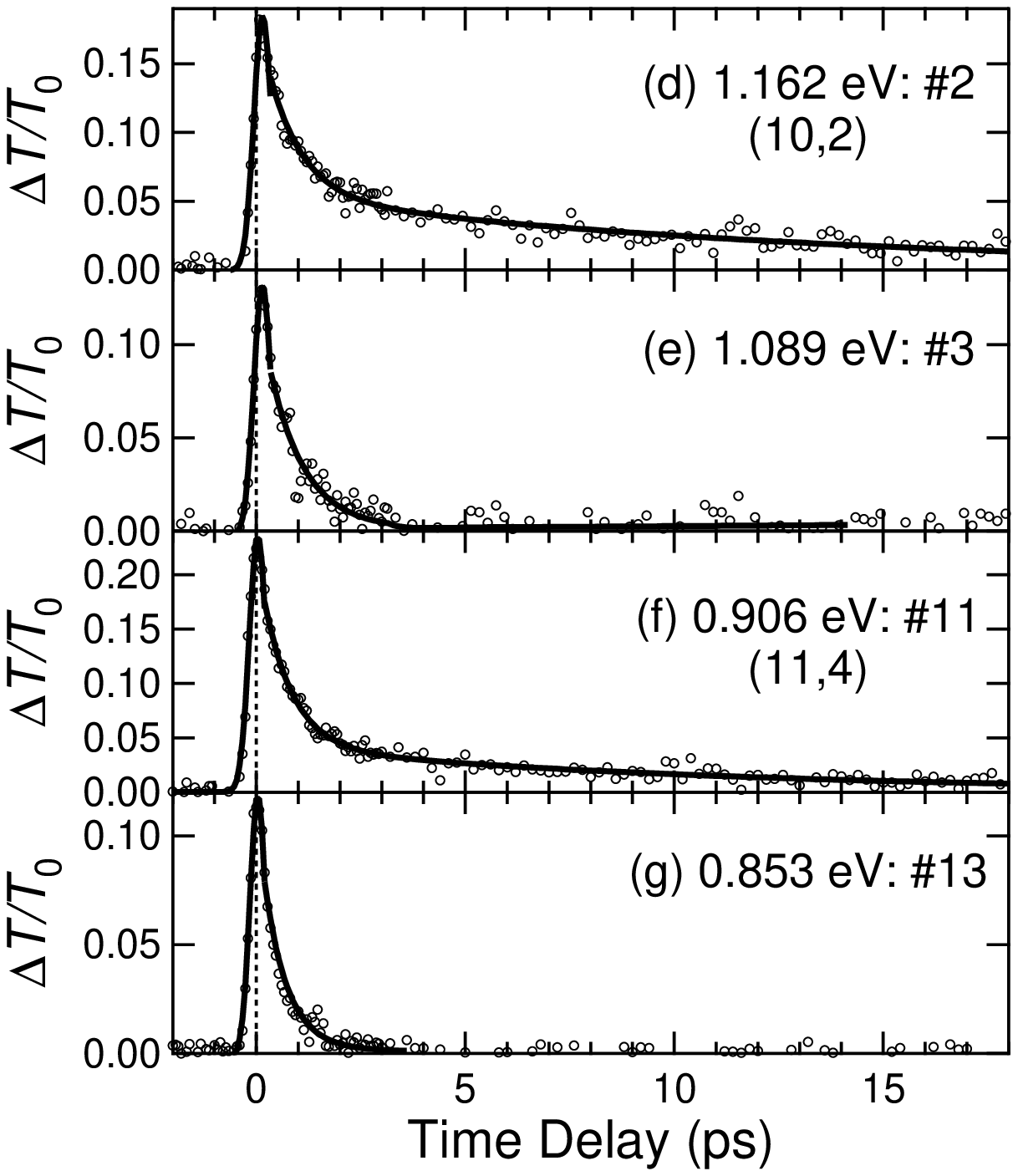}
\caption{(a) Linear absorption.  The numbers
correspond to the energies at which pump-probe measurements
were made.  (b) The peak value of $\Delta T/T_0$ and (c) the
ratio of slow to fast components vs.~photon energy.
(d)-(g): Pump-probe data taken at absorption peaks and valleys.
}
\label{summary}
\end{center}
\end{figure}

To study resonant versus non-resonant excitations
as well as chirality-dependent dynamics,
we scanned the photon energy from
0.8 to 1.1 eV, covering the E$_1$H$_1$ transitions of 0.82-1.29 nm
diameter tubes \cite{Science2,Weisman03NL}.  The
fluence was kept at $\sim$1 mJ/cm$^{2}$, below
saturation [Fig.~\ref{typical}(b)].  Note that
the probe photon energy was always the same as the pump photon energy
(degenerate pump-probe spectroscopy).
Figure \ref{summary}(a) shows absorption, which indicates the photon
energies at which we
performed pump-probe measurements (labeled 1-16).
Figure \ref{summary}(b) shows the maximum value of
$\Delta T /T_0$ vs.~photon energy; it loosely follows the absorption in (a).
Shown in Fig.~\ref{summary}(c)
is the ratio of the slow component ($\Delta T/T_0$ at 5 ps) to the fast
component ($\Delta T/T_0$ at 0 ps);
it also follows the absorption in (a), indicating
that the {\it slow component is resonantly enhanced at absorption peaks}.
Examples are shown in Figs.~\ref{summary}(d)-\ref{summary}(g).
The chosen photon energies correspond to 2, 3, 11, and 13 in (a).
The slow component is clearly observable only at absorption peaks.

\begin{figure}
\begin{center}
\includegraphics [scale=0.61] {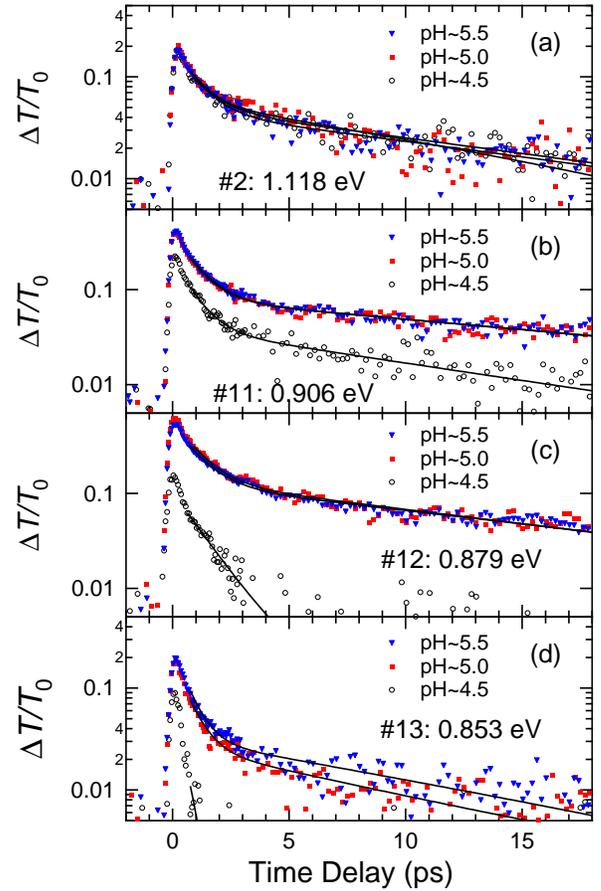}
\caption{pH-dependent pump-probe data for different wavelengths.
The pH dependence becomes stronger for smaller photon energies (or
larger-diameter tubes).}
\label{pH}
\end{center}
\end{figure}

We also found that pump-probe dynamics are strongly dependent on the pH of the
solution and the dependence is stronger at longer wavelengths (or larger
tube diameters).  Specifically, we observed that the slow component
drastically diminishes as the pH is reduced.  Examples are shown in
Fig.~\ref{pH}.  As shown in Fig.~\ref{linear}(b) and \cite{StranoetAl03JPCB},
adding hydrogen ions, H$^+$,
to the solution (or, equivalently, decreasing the pH value) diminishes,
and finally collapses, linear absorption peaks.  This effect starts from the
longer wavelength side, i.e., from larger diameter tubes.  The corresponding
reduction and disappearance of the slow component shows exactly the same trend
(see Fig.~\ref{pH}).

We believe that the key to understanding the two-component
decay is the co-existence of different types of tubes, i.e., sample
inhomogeneity.  Even when the pump
photon energy resonates with one tube type, {\em all the other tubes with
smaller band gaps are non-resonantly excited at the same time}.
Thus, the
pump-induced absorption bleaching due to state filling can relax in two
different ways: i) intraband relaxation towards the
band edge in the non-resonantly-excited tubes and ii) interband recombination
in the resonantly-excited tubes.  Process i)
is fast (as evidenced in the E$_2$H$_2$ case and in \cite{PP2,PP3,PP4})
and always exists, irrespective of the pump wavelength, unless we resonantly
excite the smallest-bandgap tubes within the diameter distribution in the sample.
In addition, it is likely that there is some coherent contribution to the pump-probe
signal in the fast component.  However, since we have not observed any
four-wave mixing signal, we do not have an estimate on the dephasing time and thus
the magnitude of this contribution.
On the other hand, Process ii) is resonantly enhanced when the pump photon energy
coincides with the absorption peak of one tube type since such
resonantly created carriers occupy states at/near the band edge and do not have
lower-energy intraband states to relax into.

There is an {\em intimate relationship between PL and the slow decay component}.
In the previous work \cite{PP2,PP3,PP4} neither PL nor the slow component
was observed whereas our sample exhibits
PL peaks at the same energies as E$_1$H$_1$ absorption peaks.  In addition,
PL and absorption peaks diminish with decreasing pH with a concomitant
abrupt disappearance of the slow decay (see Fig.~\ref{linear}, Fig.~\ref{pH},
and \cite{StranoetAl03JPCB}).
These facts lead us to believe that our pump-probe dynamics probes
{\em radiative} carrier recombination.
The slow decay times $\tau_2$ should be viewed as the lower boundary
of the possible range of the radiative lifetimes, $\tau_r$, of SWNTs.
The upper limit of $\tau_r$ depends on the rates of non-radiative recombination.
Although the ratio of radiative to non-radiative rates can in principle be estimated
from the quantum yield $\eta$, there have been no definitive studies of
$\eta$ in micelle-suspended SWNTs.
The difficulty in determining $\eta$
comes from the simultaneous occurrence of absorption by
different types of tubes for a given excitation photon energy.
Furthermore, very recent work has demonstrated that $\eta$
strongly varies with the type of surfactants used \cite{Moore}.
Hence, precise determination of $\tau_r$ requires understanding of the co-existing
non-radiative carrier recombination processes.

There are non-radiative processes that we can rule out:
first, we are not seeing any pump-power-dependent decay times, which
excludes nonlinear, non-radiative recombination, i.e.,
Auger recombination.
Second, the contact and electronic coupling between different
tubes could make the system ``gapless'';
carriers that are supposedly created in semiconducting
tubes could then recombine through purely intraband relaxation pathways
due to the existence of continuum states between the conduction and valence
bands as a result of interaction with metallic tubes.
However, tube-tube interaction is minimized in our SDS-suspended
SWNTs as evidenced by many peaks
in linear spectra (Fig.~\ref{linear}), so this can be excluded as well.
Finally, recombination on
the catalyst particles at the tube ends can be ruled out since cut tube samples
still show the same PL properties.
One possible scenario is non-radiative recombination via surface defects
in incompletely-micelle-covered nanotubes;
this is conceivable in view of the surfactant-dependent radiative
efficiency \cite{Moore}.  It should be noted that SDS shows the most prominent pH
dependence while other anionic surfactants do not show the same effect.
In other words, the nature of the surfactant plays an important role in the
surface chemistry, and thus could influence non-radiative recombination rates.
Further investigations are necessary to elucidate the origins of non-radiative
recombination.

Finally, we discuss the intriguing pH dependence.
The main effect of a decreasing pH (or an increasing H$^+$ concentration) on
pump-probe data is to destroy the slow relaxation component.  This process begins
to occur at a larger pH for a longer wavelength
(or a larger diameter tube).  We attribute this to the entrance of the Fermi
energy into the valence band.  Adsorbed H$^+$ ions act as acceptors, making the
tubes $p$-type.  When the density of such acceptors is high,
the Fermi energy can go into the valence band, destroy the interband absorption
peak, and thereby disable the mechanism for resonantly enhancing the slow relaxation
component.  This can be viewed as a 1-D manifestation of the well-known
Burstein-Moss effect \cite{BM}.
It is important to note that wider tubes have smaller effective masses,
and therefore smaller
acceptor binding energies; this explains why larger-diameter tubes are
more easily affected by a pH decrease.
 
In summary, we have observed novel two-component dynamics
in pump-probe spectroscopy experiments of SDS-suspended SWNTs.
We attribute the previously-unobserved slow component (5-20 ps)
to interband carrier recombination, which is resonantly enhanced whenever
the pump photon energy coincides with the absorption peak of one tube type.
The fast component (0.3-1.2 ps) is always present and we attribute it
to intraband relaxation in non-resonantly-excited tubes.
The slow component diminishes drastically
with decreasing pH, as a result of the Burstein-Moss effect.

We gratefully acknowledge support from the
Robert A.~Welch Foundation (Grant No.~C-1509), the Texas Advanced
Technology Program (Project No.~003604-0001-2001), and the
National Science Foundation CAREER Award (Grant No.~DMR-0134058).

 

\begin{references}

\bibitem{loudon} See, e.g., R. Loudon, Am. J. Phys. {\bf 27}, 649 (1959);
R. J. Elliot and R. Loudon, J. Phys. Chem.
Solids {\bf 8}, 382 (1959);
T. Ogawa and T. Takagahara, Phys. Rev. B
{\bf 43}, 14325 (1991);
B. Y.-K. Hu and S. Das Sarma,
Phys. Rev. Lett. {\bf 68}, 1750 (1992);
T. Ando, J. Phys. Soc. Jpn. {\bf 66},
1066 (1997).















\bibitem{dresselhaus01} {\it Carbon Nanotubes},
eds. M. S. Dresselhaus {\it et al}. (Springer, Berlin, 2001).

\bibitem{ajikiJPSJ93} H. Ajiki and T. Ando, J. Phys. Soc. Jpn.
{\bf 62}, 1255 (1993);
W. Tian and S. Datta, Phys. Rev. B {\bf 49}, 5097
(1994);
J. P. Lu, Phys. Rev. Lett. {\bf 74}, 1123 (1995).

\bibitem{alonPRL00} O. E. Alon {\it et al}.,
Phys. Rev. Lett. {\bf 85}, 5218 (2000).

\bibitem{katauraSM99} H. Kataura {\it et al}.,
Synth. Metals {\bf 103}, 2555 (1999);
S. Kazaoui {\it et al}.,
Phys. Rev. B {\bf 60}, 13339 (1999);
J. Hwang {\it et al}., Phys. Rev. B {\bf 62}, R13310 (2000).

\bibitem{Science1} M. J. O'Connell {\it et al}., Science
{\bf 297}, 593 (2002).

\bibitem{Science2}  S. M. Bachilo {\it et al}.,
Science {\bf 298}, 2361 (2002).

\bibitem{Weisman03NL}
R. B. Weisman and S. M. Bachilo, Nano Lett.
{\bf 3}, 1235 (2003).

\bibitem{PP1} T. Hertel and G. Moos,
Phys. Rev. Lett. {\bf 84}, 5002 (2000).

\bibitem{PP2} Y. C. Chen {\it et al}.,
Appl. Phys. Lett. {\bf 81}, 975 (2002).

\bibitem{PP3} H. Han {\it et al}., Appl. Phys. Lett. {\bf 82}, 1458 (2002).

\bibitem{PP4} J. S. Lauret {\it et al}., Phys. Rev. Lett.
{\bf 90}, 057404 (2003).



\bibitem{BM} E. Burstein, Phys. Rev. {\bf 93}, 632 (1954);
T. S. Moss, Proc. Phys. Soc. (London) B {\bf 67}, 775 (1954).

\bibitem{StranoetAl03JPCB}
M. S. Strano {\it et al}.,
J. Phys. Chem. B {\bf 107}, 6979 (2003).



\bibitem{Moore} V. C. Moore {\it et al}.,
Nano Lett. {\bf 3}, 1379 (2003).

\end{references}

\end{document}